\input{aipcheck.tex}
\documentclass[
final            % use final for the camera ready runs
,numberedheadings % uncomment this option for numbered sections
]
{aipproc}
\layoutstyle{6x9}

\usepackage{bm}

\SetInternalRegister\hbadness{8000} % pseudo latin isn't breaking very well :-)
\newcommand{\be}{\begin{equation}}
\newcommand{\ee}{\end{equation}}

\newcommand{\cur}{\mbox{\footnotesize cur}}
\newcommand{\dyn}{\mbox{\footnotesize dyn}}
\newcommand{\CL}{\mbox{\footnotesize CL}}
\newcommand{\mcur}{m_{\mbox{\footnotesize cur}}}

\newcommand{\mdyn}{m_{\mbox{\footnotesize dyn}}}

\newcommand{\qllsm}{QLL$\sigma$M}
\begin{document}
\title[]{Pion and Kaon Masses and Pion Form Factors  
from Dynamical Chiral-Symmetry Breaking with Light Constituent Quarks} 
\author{Michael D.\ Scadron}{
address={Physics Department, University of Arizona, Tucson, AZ 85721, USA}}
\author{Frieder Kleefeld$^1$}{
address={Centro de F\'{\i}sica das Interac\c{c}\~{o}es
Fundamentais, Instituto Superior T\'{e}cnico, Edif\'{\i}cio Ci\^{e}ncia,
P-1049-001 Lisboa, Portugal}} \footnotetext{Home address: Pfisterstr.\ 31,
D-90762 Fuerth, Germany}
\author{George Rupp}{
address={Centro de F\'{\i}sica das Interac\c{c}\~{o}es
Fundamentais, Instituto Superior T\'{e}cnico, Edif\'{\i}cio Ci\^{e}ncia,
P-1049-001 Lisboa, Portugal}}

\begin{abstract}
Light constituent quark masses and the corresponding dynamical quark masses
are determined by data, the quark-level linear $\sigma$ model, and infrared
QCD. This allows to define effective nonstrange and strange current quark
masses, which reproduce the experimental pion and kaon masses very accurately,
by simple additivity. In contrast, the usual nonstrange and strange current
quarks employed by the Particle Data Group and Chiral Perturbation Theory
do not allow a straightforward quantitative explanation of the pion and kaon
masses.  
\end{abstract}

\maketitle

\begin{center}
{\normalfont\fontsize{14}{16}\bfseries
\vskip 2mm
INTRODUCTION
}
\end{center}
\vskip 2mm
The pion is commonly accepted to be massless in the chiral limit (CL). Not
only is its physical mass a good measure of chiral-symmetry breaking (ChSB),
but also the related nonstrange constituent quark mass $\hat{m}$. In the present
short note, we shall show that $\hat{m}$ can be additively decomposed into
a bulk part called dynamical quark mass ($\mdyn$), associated with
chiral-symmetric strong interactions, and a smaller part called current quark
mass ($\mcur$), which arises from ChSB in the electroweak sector.
This \em effective \em \/$\mcur$ turns out to be precisely half the pion mass.
Moreover, the strange constituent quark mass allows a similar decompostition
as well, with the kaon mass being the simple sum of the effective nonstrange
and strange current quark masses. \\
\begin{center}
{\normalfont\fontsize{14}{16}\bfseries
QUARK-MASS DIFFERENCE
\raisebox{-2.75pt}{\includegraphics[height=14pt]{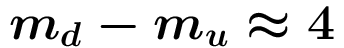}}
\hspace{-1mm}M\lowercase{e}V \\[-1mm]
}
\end{center}
\vskip 3mm
A simple estimate of the mass difference between a down and an up quark gives
%\vskip-1mm
\be
\left.\begin{array}{c}
m_{K^0}-m_{K^+} \\ m_{\Sigma^-}-m_{\Sigma^+}
\end{array} \right\}
\;\;\;\Longrightarrow\;\;\;m_d-m_u\;\approx\;4\;\mbox{MeV}\; .
\ee
\vskip 1mm

\noindent Note that this holds for both current \em and \em \/constituent
quarks. On the other hand, from the proton magnetic moment we can derive
\cite{SDR06} an average constituent quark mass as
\vskip -1.5mm
\be
\hat{m} \; = \; (m_u+m_d)/2 \; = \; 337.5 \; \mbox{MeV} \; . 
\ee
\vskip -0.5mm
Using Eq.~(1), this yields the constituent masses
\vskip -1.5mm
\be
m_u \; \approx \; 335.5 \; \mbox{MeV} \;\;\; , \;\;\;
      m_d \; \approx \; 339.5 \; \mbox{MeV} \; . 
\ee
\vskip -0.5mm
We can also obtain $\hat{m}$ in the context of the quark-level linear $\sigma$
model (\qllsm), via the Goldberger-Treiman relation \cite{SDR06,DLS99}
\vskip-2mm
\be
\hat{m} \; \approx \; f_\pi g \; = \; 93\:\mbox{MeV} \times
\frac{2\pi}{\sqrt{3}} \; \approx \; 337.4 \; \mbox{MeV} \; . 
\ee
\vskip-1mm
\noindent The agreement with the value in Eq.~(2) is remarkable. \\

\begin{center}
{\normalfont\fontsize{14}{16}\bfseries
DYNAMICAL QUARK MASS
\hspace{1mm}\raisebox{-2.75pt}{\includegraphics[height=14pt]{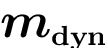}}
}
\end{center}
\vskip 2mm
A bulk dynamical CL nonstrange quark mass can be estimated as
\vskip-1mm
\be
\mdyn \; \approx \; \frac{m_N}{3} \; = \; 313 \; \mbox{MeV} \; . 
\ee
\vskip-1mm
A check via the CL pion charge radius gives
\be
\mdyn \; \approx \; \frac{\hbar c}{r_\pi^{\CL}} \; = \;
\frac{197.3\;\mbox{MeV$\cdot$fm}}{0.63\;\mbox{fm}}\; = \;313\;\mbox{MeV}\;,
\ee
employing vector-meson dominance or the \qllsm\ to predict
\vskip -1.5mm
\be
r_\pi^{\CL} \; = \; 0.63 \; \mbox{fm} \; . 
\ee
\vskip -0.5mm
\noindent Alternatively, using infrared QCD, with $\alpha_s\approx0.5$ at a
1~GeV cutoff, we get
\be
m_{\dyn} \; = \; \left[\frac{4\pi}{3}\,\alpha_s\,\langle-\bar{q}q\rangle
\right]^{\frac{1}{3}} \; \approx \; 313 \; \mbox{MeV}  \; ,
\ee
for the commonly accepted value of the quark condensate
\be
\langle-\bar{q}q\rangle \; \approx \; (245\;\mbox{MeV})^3 \; . 
\ee
\vskip 5mm
\begin{center}
{\normalfont\fontsize{14}{16}\bfseries 
EFFECTIVE CURRENT QUARK MASS VIA QCD
}
\end{center}
\vskip 1.5mm
Away from the CL, we define the effective current quark mass as
\vskip -1mm
\be
\hat{m}_{\cur} \; = \; \hat{m} - m_{\dyn} \; ,
\ee
where $\hat{m}$ is the constituent quark mass, and the dynamical mass
$m_{\dyn}$ runs as 
\vskip -1mm
\be
m_{\dyn}(p^2) \; \sim \; p^{-2} 
\ee
\noindent according to QCD.
On the $\hat{m}=337.5$ MeV mass shell, selfconsistency then requires
\be
m_{\dyn}(p^2\!=\!\hat{m}^2)\;=\;\frac{m_{\dyn}^3}{\hat{m}^2}\;=\;
\frac{(313)^3}{(337.5)^2}\;\mbox{MeV}\;=\;269.2\;\mbox{MeV} \;. 
\ee
\vskip-1mm
\noindent This yields
\be
\hat{m}_{\cur}\;=\;(337.5-269.2)\;\mbox{MeV}\;=\;68.3\;\mbox{MeV}\;,
\ee
near the pion-nucleon sigma term
\be
\sigma_{\pi N}  =  (55\pm13)\;\mbox{MeV} \; \mbox{\cite{HJS71}} \;\;,\;\;
\sigma_{\pi N}  =  (66\pm9 )\;\mbox{MeV} \; \mbox{\cite{NO74}}  \;\;,\;\;
\sigma_{\pi N}  =  (64\pm8 )\;\mbox{MeV} \; \mbox{\cite{K82}} \; . 
\ee
Note that both $\sigma_{\pi N}$ and $\hat{m}_{\cur}$ vanish in the CL.
\mbox{ } \\[1.5mm]
\begin{center}
{\normalfont\fontsize{14}{16}\bfseries 
PION 
\raisebox{-2.75pt}{\includegraphics[height=16pt]{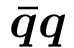}}
MASS
}
\end{center}
\vskip 2.5mm
With the effective current quark mass derived in Eq.~(13), we get a
$\bar{q}q$ pion mass 
\vskip-1mm
\be
m_\pi \; = \; 2\,\hat{m}_{\cur} \; = \; 136.6 \; \mbox{MeV} \; , 
\ee
\vskip .5mm
\noindent almost midway between the observed $m_{\pi^0}=134.98$ MeV and
$m_{\pi^+}=139.57$ MeV!  \\[1.2mm]
\begin{center}
{\normalfont\fontsize{14}{16}\bfseries 
PION FORM-FACTOR RATIO
}
\end{center}
\vskip 2.5mm
In the QLL$\sigma$M, the conserved-vector-current pion form-factor ratio
is predicted as 
\vskip -1mm
\be
\frac{F^\pi_A(0)}{F^\pi_V(0)}\; =\;1-\frac{1}{3}\;=\;\frac{2}{3}\;,
\ee
\vskip -1mm
\noindent for $q^2=0$. Here, the terms 1 and 1/3 are due to quark and meson
loops, respectively, and the relative minus sign is a Feynman rule. The
result is very near data \cite{PDG06} at
\vskip -1mm
\be
\frac{(0.0116\pm0.0016)}{(0.017\pm0.008)}\;=\;0.68 \pm 0.33 \; .
\ee
\mbox{ } \\[-1.5mm]
\begin{center}
{\normalfont\fontsize{14}{16}\bfseries 
KAON
\raisebox{-2.75pt}{\includegraphics[height=16pt]{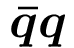}}
MASSES
}
\end{center}
\vskip 2mm
Chiral-symmetry-breaking experimental kaon masses (given $m_d-m_u\approx4$
MeV as above) are computed as (neglecting small experimental errors)
\begin{eqnarray}
m_{K^+(\bar{s}u)} \; = \; m_{s,\cur}+m_{u,\cur} \;=\; \hat{m}_{\cur}
\left[1+\left(\frac{m_s}{\hat{m}}\right)_{\!\cur}\right]-2\;\mbox{MeV}
\; = \; 493.677\;\mbox{MeV} \; , \\[2mm]
m_{K^0(\bar{s}d)} \; = \; m_{s,\cur}+m_{d,\cur} \; =\;\hat{m}_{\cur}
\left[1+\left(\frac{m_s}{\hat{m}}\right)_{\!\cur}\right]+2\;\mbox{MeV}
\; = \; 497.648\;\mbox{MeV} \; .
\end{eqnarray}
For $\hat{m}_{\cur}\approx68.3$ MeV (see Eq.~(13)), this gives \em in
both cases \em
\be
\left(\frac{m_s}{\hat{m}}\right)_{\cur}\approx6.257 \; , 
\ee
which compares well to the light-plane result \cite{SS75}
\be
\left(\frac{m_s}{\hat{m}}\right)_{\cur}\;\approx\;6\,\mbox{--}\,7 \; , 
\ee
and to other theory work \cite{GMS76}, but \em not \em \/to the
chiral-perturbation-theory (ChPT) predictions
\be
\left(\frac{m_s}{\hat{m}}\right)_{\cur}\approx25\,\mbox{--}\,30 \;\;\;
\mbox{and} \;\;\; \hat{m}_{\cur} \; \sim \; 5 \; \mbox{MeV} \; . 
\ee
\mbox{ }
\begin{center}
{\normalfont\fontsize{14}{16}\bfseries 
C\lowercase{h}PT ALTERNATIVE
}
\end{center}
\vskip 2.5mm
The value for $(m_s/\hat{m})_{\cur}$ presently adopted by the Particle Data
Group \cite{PDG06} and the ChPT \cite{GL82} research community is as large as
25--30, with the small nonstrange current-mass value
$\hat{m}_{\cur} = 2.5$--$5.5$ MeV being strongly biased by the ChPT estimate
\be
\hat{m}_{\cur} \; = \; \frac{(f_\pi m_\pi)^2}{2\,\langle-\bar{q}q\rangle}
\; \approx \; 5.5 \; \mbox{MeV} \; , 
\ee
for $f_\pi\approx93$ MeV. On the basis of these scales, the quantitative
explanation of $m_\pi$ and $m_K$ appears, however, rather cumbersome and
``unnatural'', contrary to what we observe in our aforementioned scheme
resulting from the QLL$\sigma$M and dynamically broken QCD. 

\begin{theacknowledgments}
This work was supported by the {\it Funda\c{c}\~{a}o para a Ci\^{e}ncia e a
Tecnologia} \/of the {\it Minist\'{e}rio da Ci\^{e}ncia, Tecnologia e Ensino
Superior} \/of Portugal, under contract POCI/FP/63437/2005, and by the Czech
project LC06002.

\end{theacknowledgments}

\end{document}